\documentclass[twocolumn,floatfix,prd,nofootinbib,superscriptaddress]{revtex4}

\usepackage{graphicx,epsfig}
\usepackage{amsmath}
\usepackage{amsfonts}
\usepackage{xcolor}
\usepackage{soul}

\usepackage{fancyhdr}
\usepackage{float}

\usepackage[utf8]{inputenc}
\usepackage[english]{babel}

\usepackage[export]{adjustbox}
\usepackage[colorlinks=true, allcolors=blue]{hyperref}
\usepackage{makecell}
\usepackage[justification=centering]{caption}

\usepackage{bm}
\usepackage{acro}

\DeclareAcronym{gw}{
  short = GW,
  long = gravitational wave
}
\DeclareAcronym{cw}{
  short = CW,
  long = continuous wave
}
\DeclareAcronym{ns}{
  short = NS,
  long = neutron star
}
\DeclareAcronym{mcmc}{
  short = MCMC,
  long = Markov chain Monte Carlo
}
\DeclareAcronym{lvk}{
  short = LVK,
  long = LIGO--Virgo--KAGRA
}
\DeclareAcronym{pp}{
  short = PP,
  long = percentile--percentile
}
\DeclareAcronym{pe}{
  short = PE,
  long = parameter estimation
}
\DeclareAcronym{sft}{
  short = SFT,
  long = short Fourier transform
}
\DeclareAcronym{ssb}{
  short = SSB,
  long = Solar System Barycenter
}
\DeclareAcronym{ligo}{
  short = LIGO,
  long = Laser Interferometer Gravitational-Wave Observatory
}
\DeclareAcronym{osg}{
  short = OSG,
  long = Open Science Grid
}
\DeclareAcronym{igwn}{
  short = IGWN,
  long = International Gravitational-Wave Observatory Network
}

\pdfoutput=1
\hypersetup{hidelinks} 

\allowdisplaybreaks
\newcommand{\pyfstat}{\texttt{PyFstat}}
\newcommand{\bilby}{\texttt{BILBY}}

\newcommand{\Fstat}{$\mathcal{F}$}

\newcommand{\dynesty}{\texttt{DYNESTY}}
\newcommand{\ptemcee}{\texttt{PTEMCEE}}

\begin{document}

\title{Enabling new sampling strategies for continuous-wave analyses \\ by integrating \bilby{} into \pyfstat}

\author{Maria-Antonia Ferrer-Martinez}
\email{maria-antonia.ferrer@uib.cat}
\affiliation{IAC3, Universitat de les Illes Balears, Cra. de Valldemossa km 7.5, 07122 Palma, Spain}
\affiliation{Università di Milano and INFN, sezione di Milano, Via Celoria 16, 20133 Milano, Italy}

\author{David Keitel}
\email{david.keitel@uib.cat}
\affiliation{IAC3, Universitat de les Illes Balears, Cra. de Valldemossa km 7.5, 07122 Palma, Spain}

\date{September 9, 2026}
\begin{abstract}
Continuous-wave searches for rotating neutron stars 
usually cover wide parameter spaces and produce a large number of candidate signals.
Stochastic sampling methods 
therefore play an important role in
candidate follow-up and parameter estimation. 
We present the integration of the Bayesian inference library \bilby{} into the 
open-source continuous-wave analysis package \pyfstat, 
enabling access to the broad range of stochastic samplers available through \bilby{} 
while preserving the existing \pyfstat{} analysis infrastructure. 
The implementation is validated through injection studies in Gaussian
noise for two representative single-stage follow-up configurations: candidates from a
directed search and candidates from an all-sky search.
Using the nested sampler \dynesty{}, we obtain detection efficiencies consistent 
with the theoretical sensitivity predictions.
In this setup, \dynesty{} performs comparably to the existing 
\ptemcee{}-based implementation of \pyfstat{} for the directed case,
and also provides an effective single-stage approach for the all-sky candidate follow-up 
considered here, where we did not identify a \ptemcee{} setup with similar recovery performance.
Percentile-percentile tests for \dynesty{} further show well-calibrated
credible intervals, demonstrating reliable parameter estimation.
While a full exploration of \dynesty{} settings, multi-stage setups, and other samplers
is left for future work,
these results demonstrate that this flexible and publicly available framework
opens useful new possibilities for Bayesian continuous-wave analyses.

\end{abstract}
\acresetall

\maketitle

\section{Introduction}

\Acp{gw} are expected to be emitted by non-axisymmetric rotating
\acp{ns} in the form of long-lasting, almost monochromatic \acp{cw}
\cite{rilesSearchesContinuouswaveGravitational2023}.
Although these signals have not yet been detected, they remain a key target for 
current and future \acs{gw} observatories such as \ac{lvk}, Einstein Telescope, and Cosmic Explorer
\cite{collaborationAdvancedLIGO2015, acerneseAdvancedVirgoSecondgeneration2015, akutsuOverviewKAGRADetector2021, abacScienceEinsteinTelescope2025, evansHorizonStudyCosmic2021}.
Their detection would provide unique information on the internal structure 
and dynamics of \acp{ns}, offering a direct probe of matter under 
extreme physical conditions 
\cite{jonesMultimessengerObservationsScience2025, owenMultimessengerDetectabilityContinuous2026,
Owen:2025ata}.

\Acs{cw} searches are computationally demanding, since they require the analysis 
of months to years of data over large parameter spaces. The most sensitive strategy is a 
fully coherent matched-filter search, in which the complete dataset is compared against 
a bank of signal waveform templates.
However, the computational cost of coherent searches grows rapidly with both observation 
time and parameter-space volume, making them infeasible for broad all-sky surveys, which
must cover unknown source parameters \cite{tenorioSearchMethodsContinuous2021, wetteSearchesContinuousGravitational2023}.

Most \acs{cw} searches therefore employ hierarchical strategies
\cite{bradySearchingPeriodicSources2000}.
Initial search stages use 
semicoherent methods, in which the data are divided into shorter segments that are 
analyzed coherently and then combined incoherently. This reduces the computational 
cost at the expense of some sensitivity.
Candidates identified in these wide-parameter-space searches
are subsequently followed
up with increasingly coherent analyses over progressively smaller parameter-space regions
\cite{papaHierarchicalFollowupSubthreshold2016, ashtonHierarchicalMultistageMCMC2018, tenorioApplicationHierarchicalMCMC2021, covasNewFrameworkFollow2024, mirasolaComputationallyefficientFollowupPipeline2025,  martinsBayesianFrameworkFollowup2025}.

Overall, the parameter space can be explored using either 
deterministic template grids or stochastic sampling techniques.
Grid-based methods evaluate the detection statistic
on a predefined template bank and are the standard approach for 
the initial stages of wide searches.
Stochastic methods instead explore the parameter space adaptively,
concentrating computational effort in regions favored by the data.
They are particularly well suited to constrained parameter spaces. 
This includes follow-up analyses of candidates identified in
wide-parameter-space searches
\cite{ashtonHierarchicalMultistageMCMC2018, tenorioApplicationHierarchicalMCMC2021, covasNewFrameworkFollow2024, mirasolaComputationallyefficientFollowupPipeline2025,  martinsBayesianFrameworkFollowup2025}, 
as well as targeted and narrowband searches for pulsars known from electromagnetic observations
\cite{ashokNewSearchesContinuous2021, pitkinCWInPyPythonPackage2022, ashokBayesianStatisticbasedParameter2024, collaborationSearchContinuousGravitational2025, donofrioPy5vecModularPython2026, collaborationNarrowbandSearchesContinuous2026}.

In this work, we present the integration of the Bayesian inference
library \bilby{}
\cite{ashtonBilbyUserfriendlyBayesian2019}
into the open-source \acs{cw} analysis package
\pyfstat{}
\cite{keitelPyFstatPythonPackage2021,ashtonPyFstatPyFstatV2302025}.
\pyfstat{} already provides both grid-based and stochastic search
capabilities, but its stochastic analyses have so far relied exclusively
on the parallel-tempered \ac{mcmc} sampler \ptemcee{}
\cite{foreman-mackeyEmceeMCMCHammer2013a, vousdenDynamicTemperatureSelection2016}.
Through this integration, \pyfstat{} gains access to the broad range of 
stochastic samplers provided by the \bilby{} framework,
including nested-sampling and \ac{mcmc} packages such as 
\dynesty{}
\cite{speagleDynestyDynamicNested2020},
\texttt{NESSAI}
\cite{williamsNestedSamplingNormalizing2021,williamsImportanceNestedSampling2023,michaelj.williamsMjwillNessaiV01522026},
and \texttt{EMCEE}
\cite{foreman-mackeyEmceeMCMCHammer2013a,EmceeEmcee},
among others, while retaining the existing signal model,
detection-statistic implementation, and data-analysis infrastructure
provided by \pyfstat.
Similar frameworks for $\mathcal{F}$-statistic-based analyses using
\bilby{} have recently been proposed
\cite{ashokBayesianStatisticbasedParameter2024, covasNewFrameworkFollow2024,martinsBayesianFrameworkFollowup2025} 
and have been used in open-data analyses outside the LVK collaboration
\cite{ashokNewSearchesContinuous2021, mcgloughlinEinsteinHomeAllskyBucket2025, mingObservationalConstraintsSpin2026a,Covas:2024nzs,Covas:2026raf}, 
but these implementations are not publicly available.
By integrating directly into \pyfstat, this work makes these
capabilities openly available to the broader \acs{cw} community and
enables new flexible strategies for following up candidates from the
wide variety of searches performed on \acs{lvk} data, including
Refs.~\cite{collaborationAllskySearchContinuous2025, collaborationSearchContinuousGravitational2025, collaborationAllskySearchesContinuous2026, collaborationSearchesContinuousGravitational2026, collaborationSubTorqueBalanceUpperLimits2026, collaborationNarrowbandSearchesContinuous2026}.

The remainder of this paper is organized as follows.
In Sec.~\ref{sec:background}, we introduce the \acs{cw} signal model, the
$\mathcal{F}$-statistic formalism, and the main approaches used to
explore the search parameter space.
Sec.~\ref{sec:cw-searches} reviews the stochastic sampling algorithms
considered in this work.
The implementation of \bilby{} within \pyfstat{} is described
in Sec.~\ref{sec:bilby-integration}. 
Validation studies in terms of candidate recovery and 
sampler convergence are presented in Sec.~\ref{sec:testing-framework}.
Sec.~\ref{sec:parameter-estimation} presents the parameter-estimation
results. Finally, Sec.~\ref{sec:conclusions} summarizes our conclusions. 
\section{Background}
\label{sec:background}

\subsection{Continuous-wave signal model}

The \acs{cw} signal is typically modeled as a nearly monochromatic signal with a very slowly varying frequency
\cite{jaranowskiDataAnalysisGravitationalwave1998, rilesSearchesContinuouswaveGravitational2023}.
 The strain measured at a detector can be written as
\begin{align}
h(t) &= F_{+}(t)h_{+}(t) + F_{\times}(t)h_{\times}(t) \nonumber \\
&= h_0 \left[ A_+ \cos\Phi(t) + A_\times \sin\Phi(t) \right],
\end{align}
where $F_{+}(t)$ and $F_{\times}(t)$ are the detector antenna-pattern
functions for the two \acs{gw} polarizations, 
$h_0$ is the intrinsic amplitude of the signal, 
$A_{+}$ and $A_{\times}$ are the amplitudes of the two polarizations,
and $\Phi(t)$ is the phase evolution of the signal as seen at time $t$ at the detector.

Separating the amplitude parameters from those for the phase
evolution, we can write the signal 
as
\cite{jaranowskiDataAnalysisGravitationalwave1998} 
\begin{equation}
    h(t,\mathcal{A},\lambda) 
    = \sum_{\mu=1}^4 \mathcal{A}^\mu h_\mu(t,\lambda).
\end{equation}
The amplitude parameters are
$\mathcal A=\{h_0,\cos\iota,\psi,\phi_0\}$, where 
$\iota$ is the inclination angle between the source's rotation axis and the line of sight, 
$\psi$ is the polarization angle, and $\phi_0$ is the initial phase. 

For an isolated \ac{ns}, 
the phase-evolution parameters are $\lambda=\{f,\dot{f},\ddot{f},...,\alpha,\delta\}$,
which include the \acs{gw} frequency and its time derivatives,
all evaluated at a fiducial reference time $\tau_{\mathrm{ref}}$, 
as well as the source sky position given by its right ascension and declination.

The intrinsic frequency of the signal 
can be written as a Taylor expansion around the reference time, $\tau_\mathrm{ref}$, 
\begin{equation}
f(\tau) = \left[f + \dot{f}(\tau-\tau_\mathrm{ref}) + \frac{1}{2}\ddot{f}(\tau-\tau_{\mathrm{ref}})^2 + ...\right],
\label{eq:instant-freq}
\end{equation}
where $\tau$ denotes the signal emission time in the source frame.
As generally assumed in the field, we do not explicitly consider proper motion
between the source and the \ac{ssb} \cite{Covas_2020}.
To account for the varying light-travel time 
between the detector and the \acs{ssb},
the timing model includes the Rømer delay 
together with relativistic corrections due to time dilation (the Einstein delay) 
and gravitational propagation (the Shapiro delay).

For sources in binary systems, $\lambda$ additionally includes the orbital parameters. 
The relation between the intrinsic source frequency and the frequency measured at the 
detector then includes an additional Doppler modulation arising from the orbital 
motion of the source.

\subsection{{$\mathcal{F}$}-statistic formalism}

The standard approach in \acs{cw} searches is to compare the detector data $x(t)$\footnote{$x(t)$ can refer to either single-detector data or, if available, to a multi-detector vector.} to a bank of signal templates through a detection statistic. In this framework, the data are described under two competing assumptions: the noise hypothesis, $\mathcal{H}_\mathrm{N}$, in which the data consist of noise only; and the signal hypothesis, $\mathcal{H}_\mathrm{S}$, in which the data contain both noise and a \acs{cw} signal, $h(t, \lambda, \mathcal{A})$:
\begin{align}
\mathcal{H}_\mathrm{N}: x(t) &= n(t) , \nonumber \\
\mathcal{H}_\mathrm{S}: x(t) &= n(t) + h(t, \lambda, \mathcal{A}) .
\end{align}

The corresponding likelihood ratio of the data under these two hypotheses is given by
\begin{equation}
    \mathcal{L}(x;\mathcal{A}, \lambda) = \frac{p(x|\mathcal{H}_\mathrm{S}, \mathcal{A}, \lambda)}{p(x|\mathcal{H}_\mathrm{N})} ,
\end{equation}
where $p(x|\mathcal{H}_\mathrm{S}, \mathcal{A}, \lambda)$ is the likelihood of the data under the signal hypothesis and $p(x|\mathcal{H}_\mathrm{N})$ is the likelihood of the data under the noise hypothesis.
Since $\mathcal{H}_\mathrm{S}$ is a composite hypothesis, the likelihood ratio depends on the signal parameters $\mathcal{A}$ and $\lambda$.
The amplitude parameters $\mathcal{A}$ can be handled by either maximization or marginalization
\cite{prixTargetedSearchContinuous2009},
 while the phase-evolution parameters $\lambda$ are generally searched over, except in fully targeted searches where they can be constrained by electromagnetic observations.

In the frequentist approach, the $\mathcal{F}$-statistic 
\cite{jaranowskiDataAnalysisGravitationalwave1998a, cutlerGeneralizedStatisticMultiple2005} 
is computed by maximizing the likelihood ratio over the amplitude parameters:
\begin{equation}
    \mathcal{F}(x, \lambda) = \max_{\mathcal{A}} \log \mathcal{L}(x; \mathcal{A}, \lambda) .
    \label{eq:max-likelihood}
\end{equation}

Although originally derived as a maximum-likelihood statistic, 
the $\mathcal{F}$-statistic also admits a Bayesian interpretation. 
For a particular choice of prior on the amplitude parameters, 
the $\mathcal{F}$-statistic is proportional to the logarithm of the Bayes factor
between the signal and noise hypotheses after marginalization over 
the amplitude parameters \cite{prixTargetedSearchContinuous2009}. 
This connection motivates the use of the $\mathcal{F}$-statistic 
as a detection statistic in both frequentist and Bayesian analyses.

Under ideal Gaussian-noise assumptions, the statistical distribution of the
$\mathcal{F}$-statistic is known analytically. In the absence of a signal, the
detection statistic $2\mathcal{F}$ follows a central $\chi^2$ distribution with
four degrees of freedom, 
$2\mathcal{F} \sim \chi^2_4$.
In the presence of a signal whose phase-evolution parameters match the searched
template, the distribution becomes a non-central $\chi^2$ distribution, 
$2\mathcal{F} \sim \chi^2_4(\rho^2)$,
where $\rho^2$ is the optimal signal-to-noise ratio squared. 
See Appendix~\ref{sec:appendix-theoretical-estimate} 
for a more detailed definition of $\rho^2$
and other related quantities.

\section{Stochastic sampling methods for \NoCaseChange{\acs{cw}} searches}
\label{sec:cw-searches}

One of the main applications of \pyfstat{} 
\cite{keitelPyFstatPythonPackage2021}
is the follow-up of candidates within hierarchical searches using stochastic sampling. 
These candidates may originate from a directed search, 
for which the sky position can remain fixed during the follow-up, 
or from an all-sky search,
which additionally requires exploring an uncertainty region in sky position.
Ashton and Prix
\cite{ashtonHierarchicalMultistageMCMC2018}
originally developed the package and demonstrated a hierarchical multistage \acs{mcmc} follow-up scheme in which the
coherence time is increased between stages. 
This approach was
first applied to candidates from an all-sky search by Covas and Sintes \cite{Covas:2020nwy}.
It was subsequently used to follow-up candidates identified in \acs{ligo} data by Tenorio et al.
\cite{tenorioApplicationHierarchicalMCMC2021}, as well as in various \acs{lvk} results papers from the O3 and O4 observing runs
\cite{LIGOScientific:2021ozr, LIGOScientific:2021quq, KAGRA:2022osp, LIGOScientific:2022enz, Whelan:2023bha,LIGOScientific:2025ouy,LIGOScientific:2026plm,LIGOScientific:2026ppd}.
More recently, Mirasola and Tenorio
\cite{mirasolaComputationallyefficientFollowupPipeline2025}
optimized the sampler settings and the hierarchical scheme 
for more efficient follow-ups of candidates from broad
blind searches.

In general, most stochastic sampling methods used in \acs{gw} astronomy
fall into two broad categories: \acs{mcmc}
\cite{hastingsMonteCarloSampling1970}
and nested sampling 
\cite{skillingNestedSamplingGeneral2006,ashtonNestedSamplingPhysical2022}. 
\Acs{mcmc} methods produce samples from the posterior distribution,
\begin{equation}
    p(\theta|x) \propto \mathcal{L}(x|\theta)\pi(\theta) ,
\end{equation}
where $\theta$ are the parameters of interest, $x$ are the data, $\mathcal{L}(x|\theta)$ 
is the likelihood function, and $\pi(\theta)$ is the prior distribution. 

Nested sampling, by contrast, was primarily developed to compute
the Bayesian evidence,
\begin{equation}
    Z=\int \mathcal{L}(x|\theta)\pi(\theta)\,d\theta,
\end{equation}
while also producing posterior samples.
Instead of sampling directly from the posterior distribution, 
nested sampling begins by drawing samples from the prior and 
progressively restricts the explored region to higher likelihood values. 
As the corresponding prior volume shrinks, both the Bayesian evidence and the posterior distribution can be estimated.

Several stochastic samplers are available through the \bilby{}
inference library \cite{ashtonBilbyUserfriendlyBayesian2019}, 
widely used in \acs{gw} astronomy.
It provides a common interface that separates 
the likelihood implementation from the sampling algorithm.
This design allows the same likelihood function to be explored 
using different stochastic samplers 
without modifying the existing analysis pipeline.
The integration presented in this work extends the
stochastic inference capabilities of \pyfstat, which were
previously limited to the \ptemcee{} \acs{mcmc} sampler, by making the
broad range of \acs{mcmc} and nested-sampling algorithms supported by
\bilby{} available within the existing \pyfstat{} analysis
framework. 

The following subsections briefly summarize the two samplers 
considered as examples in this work, \ptemcee{} and \dynesty{}.

\subsection{The \ptemcee{} sampler}

The \ptemcee{} sampler
\cite{vousdenDynamicTemperatureSelection2016}
is built on top of the affine-invariant ensemble \acs{mcmc} sampler \texttt{EMCEE}
\cite{foreman-mackeyEmceeMCMCHammer2013a}
and combines it with parallel tempering to improve the exploration of complex posterior distributions.

Affine invariance makes the sampler insensitive to linear rescalings, 
rotations, and translations of the parameter space.
As a result, its performance is less sensitive to 
strong parameter correlations or widely different parameter scales.

Parallel tempering evolves multiple ensembles at different temperatures.
Each ensemble samples a tempered posterior,
\begin{equation}
    p_T(\theta | x) \propto
    \mathcal{L}(x|\theta)^{1/T}\pi(\theta),
\end{equation}
where $T=1$ corresponds to the target posterior, whereas higher
temperatures flatten the likelihood surface and permit broader
exploration of the parameter space. Occasional exchanges between
temperatures help the sampler escape local maxima and improve
convergence.

The principal tuning parameters of \ptemcee{} are the number of walkers, the number 
and spacing of temperatures, and the number of sampling steps. 
Their optimal values depend on the dimensionality and complexity of the parameter space and have been discussed extensively in
Ref.~\cite{mirasolaComputationallyefficientFollowupPipeline2025}.
The configuration used in this work is described in Sec.~\ref{sec:testing-framework}.

\subsection{The \dynesty{} sampler}

\dynesty{} \cite{speagleDynestyDynamicNested2020} is an implementation of 
the nested sampling algorithm for computing the Bayesian evidence 
and producing posterior samples. 
Nested sampling maintains a set of live points, 
which collectively define the region of parameter space currently being explored.
During the run, the lowest-likelihood live
point is repeatedly replaced by a new sample with higher likelihood, 
progressively concentrating the live points towards regions of increasing likelihood.
Optionally, \dynesty{} also allows for a dynamic nested sampling mode
that extends this procedure by adapting the number of
live points throughout the run to allocate computational effort more
efficiently; this feature, however, is not used in the present work.

The principal tuning parameter of \dynesty{} is the number of live points, 
which controls the resolution with which the constrained prior is explored.
Additional tuning parameters include the sampling strategy 
used to generate new live points within the constrained prior, 
the method used to bound the sampled region, 
and the stopping criterion for the evidence calculation. 
See Ref.~\cite{ashtonNestedSamplingPhysical2022} for an accessible overview of these parameters and their impact on the sampling performance.
The specific choices adopted in this work are described in Sec.~\ref{sec:testing-framework}. 

\section{Integration of \bilby{} into \pyfstat}
\label{sec:bilby-integration}
\subsection{Software interface}

The integration\footnote{This implementation is not yet available in released versions or on the \pyfstat{} master branch, but is already publicly available while undergoing code review before merging.}
 of \bilby{} into \pyfstat{} builds on the
existing \pyfstat{} \acs{mcmc} classes
\cite{keitelPyFstatPythonPackage2021,ashtonPyFstatPyFstatV2302025}.
A new \texttt{run\_bilby} method is introduced in the
\texttt{MCMCSearch} class, providing an alternative entry point for
running an otherwise standard \pyfstat{} analysis. 
The signal model, input data, sampled and fixed parameters, and 
detection-statistic computation are handled through \pyfstat{},
while stochastic sampling is delegated to \bilby{}. 
Prior distributions specified via \pyfstat{}
are converted into a \bilby{} prior dictionary,
although priors defined directly through \bilby{} may also be
provided. The selected sampler and its configuration are then passed 
to the standard \texttt{bilby.run\_sampler} interface. 
Consequently, any sampler supported by \bilby{} can be used 
while preserving the existing \pyfstat{} classes,
\Fstat-statistic calculations, and post-processing tools.

To connect the two packages, we introduce the
\texttt{PyFstatBilbyLikelihood} class, which implements the likelihood
interface expected by \bilby{} while delegating the
detection-statistic calculation to \pyfstat{}.
This design allows the \bilby{} integration to be used with classes
derived from \texttt{MCMCSearch} without requiring separate likelihood
implementations. 
Such classes in \pyfstat{} currently include, e.g., 
fully coherent, semicoherent, glitch-robust
\cite{ashtonSemicoherentGlitchrobustContinuous2018},
and transient searches 
\cite{keitelFasterSearchLong2018}.
For each parameter-space point proposed by the sampler, 
the wrapper translates the sampled parameters into the 
ordering expected by the corresponding \pyfstat{} class 
and delegates the \Fstat-statistic evaluation to the
existing search implementation. 

Once \texttt{bilby.run\_sampler} completes the sampling run, 
the returned posterior samples and likelihood values are mapped 
onto the standard \pyfstat{} result structures while retaining 
the original \bilby{} result object. 
This allows users to combine both the \bilby{} result-analysis tools 
and the existing \pyfstat{} post-processing functionality, 
including sample export, posterior plots, search summaries, 
and loudest-candidate generation.

The current implementation places the \texttt{run\_bilby} method 
within the existing \texttt{MCMCSearch} class to preserve compatibility 
with the current \pyfstat{} architecture. 
Since \bilby{} supports both \acs{mcmc} and nested-sampling methods, 
the class name is not fully representative of the available sampling methods.
In future developments, the class structure may be refactored to
better reflect the broader scope of the \bilby{} integration.

\subsection{Likelihood formulation}

The likelihood supplied to \bilby{} is evaluated at each set of
phase-evolution parameters sampled by the corresponding \pyfstat{} search class.
For each such point, \pyfstat{} computes the $\mathcal{F}$-statistic, which
analytically maximizes the likelihood over the amplitude parameters
(see Sec.~\ref{sec:background}).
Following the Bayesian interpretation of the $\mathcal{F}$-statistic presented in
Ref.~\cite{prixTargetedSearchContinuous2009}, marginalization over the amplitude
parameters under the canonical amplitude prior yields a signal-to-noise Bayes
factor that can be expressed directly in terms of the $\mathcal{F}$-statistic.
The corresponding signal log-likelihood used by \bilby{} is therefore
written as
\begin{equation}
\log \mathcal{L}_\mathrm{{S}}(x,\lambda)
=
\mathcal{F}(x,\lambda)
+
\log\left(\frac{70}{\hat{\rho}_{\max}^4}\right)
+
\log \mathcal{L}_\mathrm{N}(x),
\label{eq:logL}
\end{equation}
where $\hat{\rho}_{\max}$ denotes 
the upper cutoff 
on the signal-strength prior introduced in
Ref.~\cite{prixTargetedSearchContinuous2009}. 
Here, $\mathcal L_\mathrm N$ is the 
Gaussian-noise likelihood,
\begin{equation}
    \mathcal{L}_\mathrm{N}(x)
    = P({x}|\mathcal{H}_\mathrm{N})
    = \kappa\exp\left[-\frac{1}{2}({x}|{x})\right],
\label{eq:L_noise}
\end{equation}
where $(x|x)$ denotes the noise-weighted inner product of the data $x$ with themselves and $\kappa$ is a normalization constant. 
The full explicit form of $\mathcal{L}_\mathrm{N}$ is derived in
Appendix~\ref{sec:noise-likelihood}.

For a fixed dataset, both $\log \mathcal{L}_\mathrm{N}$ and
$\log(70/\hat{\rho}_{\max}^4)$ are independent of
$\lambda$ and therefore contribute only additive constants
to the signal likelihood in Eq.~\ref{eq:logL}.
Consequently, for a fixed dataset, the posterior distribution over
the sampled parameters is
\begin{equation}
p(\lambda|x,\mathcal{H}_{\mathrm{S}})
\propto
\exp\left[\mathcal{F}(x,\lambda)\right]
\pi(\lambda|\mathcal{H}_{\mathrm{S}}),
\end{equation}
where $\pi(\lambda|\mathcal{H}_{\mathrm{S}})$ denotes the prior on the sampled
phase-evolution parameters.

Previous stochastic sampling implementations in \pyfstat{}
also followed the Bayesian interpretation of the $\mathcal{F}$-statistic
described in Ref.~\cite{prixTargetedSearchContinuous2009},
including the normalization term
$\log(70/\hat{\rho}_{\max}^4)$ arising from the canonical amplitude prior,
but omitted the Gaussian-noise likelihood.
In the present implementation, this term is additionally retained to provide
a consistently normalized likelihood necessary for the evidence calculations performed
by \bilby{} and, in particular, for evaluating Bayes factors between
the signal and Gaussian-noise hypotheses.

\section{Testing the new framework}
\label{sec:testing-framework}

To validate the new implementation, we choose the \dynesty{}
sampler as an example, since it is widely used
for Bayesian inference across \ac{gw} astronomy
\cite{10.1093/mnras/staa2850,LIGOScientific:2026ifv},
including in \ac{cw} analyses
\cite{pitkinCWInPyPythonPackage2022,covasNewFrameworkFollow2024,martinsBayesianFrameworkFollowup2025}.
We then perform a set of controlled injection studies in simulated
Gaussian noise.
The primary goal is to verify that the
measured sensitivity is consistent with the expected behavior
of fully coherent \ac{cw} analyses, 
rather than to optimize sampler configurations for specific
realistic applications.

We characterize the sensitivity through the detection probability,
$p_\mathrm{det}$, at fixed false-alarm probability, $p_\mathrm{fa}$, for
signal populations spanning a range of amplitudes, 
following the methodology of previous \ac{cw} sensitivity studies
\cite{shaltevFullyCoherentFollowup2013, ashtonHierarchicalMultistageMCMC2018}.
The signal amplitude relative to the detector noise level is expressed in terms of the
sensitivity depth, $\mathcal{D}$
\cite{behnkePostprocessingMethodsUsed2015,dreissigackerFastAccurateSensitivity2018, wetteSearchesContinuousGravitational2023},
defined as 
\begin{equation}
    \mathcal{D} = \frac{\sqrt{S_n}}{h_0} ,
\end{equation}
where $S_n$ is the one-sided noise power spectral density and $h_0$ 
is the intrinsic \ac{gw} strain amplitude.

We additionally compare the performance of our reference 
\dynesty{} sampler with that
of the existing \ptemcee{} implementation in \pyfstat.
This comparison provides a reference for assessing both the
sensitivity and the computational cost of the new implementation.

\subsection{Setup}
\label{sec:setup}

We assess the performance of the \bilby{}-based integration of
\dynesty{} within \pyfstat{} in two representative \ac{cw} follow-up
scenarios: the follow-up of candidates from a directed search and from
an all-sky search.
For the directed search follow-up, the sky position of the source is
assumed to be known, and the analysis is performed only over frequency
and spin-down parameters (see Sec.~\ref{sec:directed}).
For the all-sky search follow-up, the sky position is also included
among the sampled parameters (see Sec.~\ref{sec:allsky}).
$\ddot f$ and higher-order spin-down terms
are held fixed in the analyses presented here, 
although the interface also supports searches over these parameters\footnote{Full support for $\dddot f$ and higher-order spin-downs is available is available in the latest \pyfstat{} release 2.4.0.}
and, 
for sources in binary systems, over the relevant orbital parameters.

For both follow-up configurations, 
we analyze contiguous 100-day datasets 
(similar to the benchmark used in Ref.~\cite{ashtonHierarchicalMultistageMCMC2018})
consisting of independent
realizations of Gaussian noise characterized by a fixed power spectral density.
Injected signals have
$f=30.0\,\mathrm{Hz}$ and $\dot f=-1.0\times10^{-10}\,\mathrm{Hz\,s^{-1}}$. 
The sky position is drawn from isotropic distributions with 
$\alpha\sim\mathcal{U}(0,2\pi)$ and $\sin\delta\sim\mathcal{U}(-1,1)$. 

The source orientation parameters are also sampled isotropically, 
with $\cos\iota\sim\mathcal{U}(-1,1)$ and 
$\psi\sim\mathcal{U}(-\pi/4,\pi/4)$, 
while the initial phase is fixed to $\phi_0=0$. 
The noise amplitude spectral density is set to $\sqrt{S_n}=10^{-23}\,\mathrm{Hz}^{-1/2}$
and the signal amplitude is determined by the chosen sensitivity depth $\mathcal{D}$.

For the directed search follow-up,
uniform priors are placed on $f$ and $\dot f$ with full widths
$\Delta f = 2\times10^{-5}\,\mathrm{Hz}$ and 
$\Delta \dot f = 1.8\times10^{-10}\,\mathrm{Hz\,s^{-1}}$ 
centered on the injected signal parameters. 
For the all-sky follow-up, the search additionally includes the
sky-position parameters, using uniform priors with widths
$\Delta\alpha=\Delta\delta=0.05\,\mathrm{rad}$.

We base our search setup on the configuration used by Ashton and Prix
\cite{ashtonHierarchicalMultistageMCMC2018}, adopting the same 100-day
observation span, frequency
and sky prior widths.
We depart from that configuration in two respects. 
First, we use a spin-down width of
a factor of ten broader than the width adopted in 
Ref.~\cite{ashtonHierarchicalMultistageMCMC2018}.
Second, whereas that study considers data from a single detector, 
we use a two-detector \acs{ligo} network, as commonly employed in \acs{cw} searches. 
Our configuration is therefore not a direct reproduction 
of that study, but a more
demanding test with a larger prior volume and a multi-detector likelihood.

For the two different follow-up scenarios, we cover ranges of
sensitivity depths with sets of simulated signal injections 
(generated with the \texttt{Writer} class from \pyfstat{}, 
which wraps \texttt{lalpulsar\_Makefakedata\_v5})
\cite{lalsuite,swiglal}. 
The number of injections per depth ranges from 30 to 120,
chosen to ensure a well-constrained detection efficiency curve.

Each injection is then analyzed over the prior widths described above.
The detection statistic for each injection is taken to be 
the largest value of the fully coherent 
$2\mathcal{F}$-statistic encountered
during the nested-sampling exploration. For
consistency with standard \ac{cw} sensitivity studies, detection
efficiencies are evaluated using this maximum $2\mathcal{F}$ value rather than
the Bayesian evidence.

To determine whether a signal is recovered,
we first estimate the distribution of this statistic from a set of noise-only realizations. 
For  $p_\mathrm{fa}=0.01$, 
the threshold $2\mathcal{F}_\mathrm{th}$ is defined as the 
99th percentile of this noise-only distribution.
An injection is considered detected if the maximum sampled value satisfies
$2\mathcal{F}_{\rm max}>2\mathcal{F}_{\rm th}$,
and missed otherwise.
The detection efficiency at a given
sensitivity depth $\mathcal{D}$ is then defined 
as the fraction of injections recovered according to this criterion. 

Unless otherwise stated, the analyses use the \dynesty{} nested sampler
with the \texttt{rwalk} sampling scheme.
The directed follow-up employs 450 live points,
while the all-sky follow-up uses 1500 live points 
to account for the increased dimensionality of the search space. 
The choice of the number of live points is discussed in
Appendix~\ref{sec:appendix-nlive-values}.
Parameter-space bounds are constructed using the
\texttt{live} method, and the stopping criterion is set by an
evidence tolerance of \texttt{dlogz}=0.1;
see Ref.~\cite{speagleDynestyDynamicNested2020} for more details on these options.

To quantify the relative sizes of the two parameter spaces, we compute
the metric-based effective number of templates that would be required to cover the
parameter space at a mismatch of unity, $\mathcal{N}^*$,
following Eq.~(24) of Ref.~\cite{ashtonHierarchicalMultistageMCMC2018}. 
Following that work, we evaluate
$\mathcal{N}^*$ for an equatorial sky patch centered at
$\delta=0$, which maximizes the solid angle
subtended by fixed widths $\Delta\alpha$ and $\Delta\delta$, while
we choose $\alpha=0$ as a reproducible reference point.
Using the 100-day observation span and the prior widths given above, we obtain
$\mathcal{N}^*\approx9.9\times10^5$ for the directed follow-up and
$\mathcal{N}^*\approx2.3\times10^{10}$ for the all-sky follow-up,
corresponding to an increase in effective parameter-space size by a
factor of approximately $2.3\times10^4$.
This illustrates the substantially larger effective parameter-space volume of the all-sky follow-up, consistent with the increased sampling difficulty observed in the following sections.

In addition to the analyses of the contiguous 100-day datasets presented in this section, 
we repeat the validation study using O4a-timestamp datasets, constructed from
detector timestamps from the first part of the fourth \acs{lvk} observing run (O4a).
These datasets provide a more realistic observing schedule including gaps in the data.
The corresponding results are presented for comparison in
Appendix~\ref{app:o4timestamps}.

\subsection{Theoretical sensitivity estimates}
\label{sec:theoretical-estimates}
The injection studies provide an empirical estimate of the recovery fraction as
a function of sensitivity depth. 
To interpret these results, we compare the measured efficiencies with theoretical 
predictions for an ideal fully coherent search. 
This comparison allows us to distinguish
between missed detections caused by the signal being
too weak relative to the noise to exceed the detection threshold 
and those arising
because the stochastic sampler does not identify the highest likelihood region. 
The theoretical predictions describe the first effect, while deviations of the 
measured efficiencies from these curves 
can indicate losses due to the sampler or other limitations of the
adopted analysis configuration.

We consider two theoretical sensitivity estimates. Both are based on the known
distribution of the $\mathcal{F}$-statistic in Gaussian noise, but differ in how
the distribution of signal-to-noise ratios is modeled across the source population. 
The first is a simplified semi-analytic estimate
\cite{wetteEstimatingSensitivityWideparameterspace2012, ashtonHierarchicalMultistageMCMC2018}
that uses
population-averaged antenna-pattern factors while retaining the explicit
dependence on source inclination. The second is computed with the open-source
package \texttt{cows3} \cite{mirasolaComputationallyefficientFollowupPipeline2025},
which implements the sensitivity-estimation framework of
Ref.~\cite{dreissigackerFastAccurateSensitivity2018} and 
performs a more complete averaging over source and detector parameters
than the first semi-analytic estimate.

Both estimates use the same detection threshold obtained from the corresponding
noise-only simulations. Details of these calculations are given in
Appendix~\ref{sec:appendix-theoretical-estimate}.

\subsection{Results}
\label{sec:results}

This section presents the results of the injection studies introduced in
Sec.~\ref{sec:setup}. We first consider the directed search follow-up
and then the all-sky follow-up, comparing the measured detection efficiencies 
with the theoretical sensitivity estimates and,
in the directed case, with the existing \ptemcee{} implementation.

\subsubsection{Follow-up of candidates from a directed search}
\label{sec:directed}

In the directed case, the follow-up is restricted to frequency and spin-down 
parameters, using the priors and the injected parameters described in Sec.~\ref{sec:setup}. 

To determine the detection threshold, we analyze 1000 noise-only realizations
for both \dynesty{} and \ptemcee{}
and estimate the distribution of the maximum $2\mathcal{F}$ statistic. For $p_\mathrm{fa}=0.01$, this yields approximately the same threshold for both samplers,
$2\mathcal{F}_\mathrm{th}\approx47.4$,
which is used to classify each
injection as detected or missed. 
Because the theoretical sensitivity curves depend on
the sampler only through this threshold, 
the theoretical curves for the two samplers are identical.

\begin{figure}[tbp]
    \centering
    \includegraphics[width=\columnwidth]{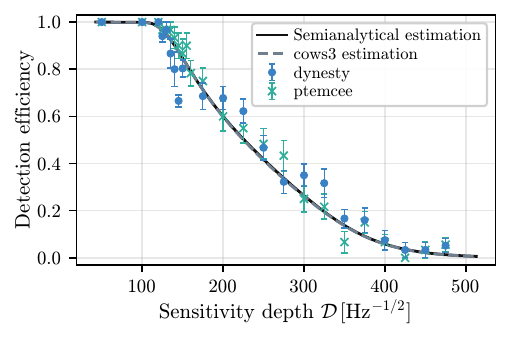}
    \caption{Detection efficiency as a function of sensitivity depth for the
    directed follow-up. Markers show the measured efficiencies obtained
    with \dynesty{} and \ptemcee{} using the sampler settings described in the text;
    the overlapping solid and dashed curves show 
    the theoretical sensitivity estimates.}
    \label{fig:directed_efficiency}
\end{figure}

Fig.~\ref{fig:directed_efficiency} shows the measured detection efficiency as
a function of sensitivity depth for the directed search follow-up, together
with the semi-analytic prediction and the \texttt{cows3} sensitivity estimate.
The detection efficiencies obtained with \dynesty{}
(\texttt{nlive}=450) closely follow both predictions over the full range of
depths explored. 
Thus, within the statistical uncertainties of the injection
study, the nested-sampling implementation recovers the expected fully coherent
sensitivity.

The corresponding detection efficiencies obtained with
\ptemcee{}
are statistically consistent with the \dynesty{} results. 
For this comparison, we used 300 walkers, 3 temperatures, 
1000 burn-in steps, and 1000 production steps.
This configuration is more expensive than those adopted
in Refs.~\cite{ashtonHierarchicalMultistageMCMC2018,mirasolaComputationallyefficientFollowupPipeline2025}, 
while cheaper configurations tested here did not achieve 
comparable performance.
A key difference explaining this is that our analysis 
performs the fully coherent follow-up
in a single stage, whereas the approaches in those references use hierarchical
multi-stage follow-ups to progressively refine the parameter space.

The agreement with both the theoretical
predictions and the existing \ptemcee{} 
validates the \bilby{}-based implementation for follow-ups of candidates from directed searches.
The results also demonstrate that a single-stage fully coherent follow-up
is feasible for this scenario using \dynesty{} with a relatively low
number of live points.

\subsubsection{Follow-up of candidates from an all-sky search}
\label{sec:allsky}
In the all-sky follow-up,
the analysis is performed over frequency, spin-down, right ascension, and declination, 
using the injected parameters and priors as described in Sec.~\ref{sec:setup}. 

To characterize the noise background, we again analyze 1000 noise-only
realizations and estimate the distribution of the maximum
$2\mathcal{F}$ statistic. 
For $p_\mathrm{fa}=0.01$, this yields $2\mathcal{F}_\mathrm{th}\approx63.4$. 
This threshold is higher than that
obtained for the fixed-sky follow-up, as expected from the larger
parameter space and the corresponding increase in the effective trials
factor.

\begin{figure}[tbp]
    \centering
    \includegraphics[width=\columnwidth]{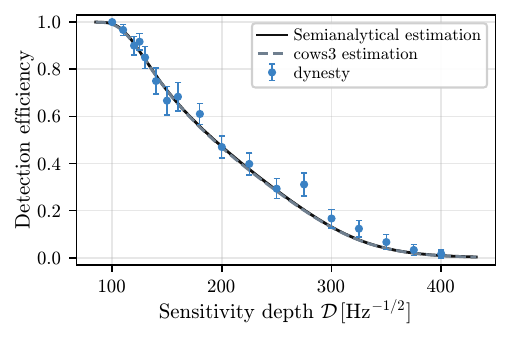}
    \caption{Detection efficiency as a function of sensitivity depth for the
    all-sky follow-up. 
    Markers show the measured efficiencies obtained with
    \dynesty{} using \texttt{nlive}=1500; the overlapping solid and dashed curves show the theoretical sensitivity estimates.
    }
    \label{fig:allsky_efficiency}
\end{figure}

The corresponding detection-efficiency results for the all-sky follow-up are
shown in Fig.~\ref{fig:allsky_efficiency}. 
The measured efficiencies remain consistent with 
both the semi-analytic prediction and the \texttt{cows3}
sensitivity estimate over the full range of depths explored, 
showing that the expected fully coherent sensitivity is 
recovered despite the substantially larger parameter space.

However, recovering the theoretical sensitivity in the larger parameter space requires
a larger number of live points
(\texttt{nlive}=1500) than in the directed search follow-up, reflecting the
increased complexity of the parameter space. 
The choice of \texttt{nlive} is motivated in Appendix~\ref{sec:appendix-nlive-values}, 
where the dependence of the recovery efficiency on the number of live points is investigated.

In contrast to the directed search follow-up,
we did not identify a single-stage
\ptemcee{} configuration that provided similarly robust performance 
for the all-sky follow-up considered here.
This does not imply that no such configuration exists.
In particular, previous work has 
shown that \ptemcee{} can be successfully applied in hierarchical multistage 
follow-up schemes, where the parameter space is progressively narrowed
\cite{ashtonHierarchicalMultistageMCMC2018,tenorioApplicationHierarchicalMCMC2021,
mirasolaComputationallyefficientFollowupPipeline2025}. 
The present comparison is instead
restricted to a single-stage search over the initial prior volume.

These results indicate that nested sampling can be successfully applied
to a single-stage fully coherent follow-up of candidates from all-sky searches,
recovering the expected sensitivity while requiring comparatively little
sampler-specific tuning for the configurations considered here.
However, it is significantly more computationally expensive
(see Sec. \ref{sec:comp-cost}) 
than the optimized hierarchical multi-stage follow-up approach from
Ref.~\cite{mirasolaComputationallyefficientFollowupPipeline2025}.

\subsection{Convergence studies}
\label{sec:convergence}

Assessing convergence in nested sampling is less straightforward 
than for \ac{mcmc} methods, 
since the algorithm terminates according to an evidence-based stopping criterion 
rather than fixed iteration counts, correlation-length estimates, or similar
diagnostics commonly used for \ac{mcmc} methods.
Although this stopping criterion indicates that the evidence estimate has stabilized, 
it does not by itself guarantee that all relevant high-likelihood regions 
have been explored. 
We therefore assess convergence empirically by means of a repeatability study of 
the reference directed search follow-up configuration adopted in this work.

For this, we fix one set of physical signal parameters, including the phase
parameters, sky position, source orientation, and sensitivity depth
$\mathcal{D}=145 \, \mathrm{Hz}^{-1/2}$, and inject this same signal into 50 independent
Gaussian-noise realizations. Each dataset is then analyzed five times,
changing only the \dynesty{} random seed.
Within each five-run group, both the injected signal and 
the noise realization are identical, 
so any differences between the recovered results isolate 
the stochastic variation introduced by the sampler. 
Across the 50 groups, we test whether this behavior is robust to the 
particular noise realization.

For 49 of the 50 noise realizations, all five searches yield 
the same binary detection decision at the threshold
$2\mathcal{F}_{\rm th}\approx47.4$, meaning that the five values are either
all above or all below the threshold; the values themselves need not be identical.
To quantify the numerical agreement, we compute, for each dataset, the
difference between the largest and smallest recovered
$2\mathcal{F}_{\rm max}$ values across its five sampler repeats. The
median of this difference over the 50 datasets, including the anomalous case
described below, is only $0.004$, demonstrating that the recovered
detection statistic is highly reproducible.

Among the 250 analyses, a single sampler realization fails to recover 
the high-likelihood region associated with the injected signal. 
For one of the 50 noise realizations, four runs recover a $2\mathcal{F}_{\rm max}\approx49.3$, 
whereas the remaining run returns $2\mathcal{F}_{\rm max}\approx35.8$. 
Its natural-log evidence, $\ln Z$, is also lower by approximately $4.8$
natural-log units
relative to the median of the other four runs. 
Since the injected signal and noise realization are identical in all five repeats, 
this discrepancy can only be attributed to the stochastic behavior of the sampler.

We next examine the repeatability of the Bayesian evidence estimates.
Let $\ln Z_{ij}$ and $\sigma_{\ln Z,ij}$ denote the natural-log evidence estimate
and its reported uncertainty for repeat $j$ of dataset $i$, and
$s_i(\ln Z)$ be the sample standard deviation of the five
recovered evidence estimates for dataset $i$. 
We quantify the agreement between the observed scatter 
and the reported uncertainty using
\begin{equation}
R_{Z,i} =
\frac{s_i(\ln Z)}
     {\frac{1}{5}\sum_{j=1}^{5}\sigma_{\ln Z,ij}}.
\end{equation}
For a set of five independent Gaussian estimates, the theoretical
median of the measured standard deviation is approximately $0.92$
times the underlying standard deviation
~\cite[Theorem~5.3.1(c)]{CasellaBerger2002}.
Individual values of $R_{Z,i}$ fluctuate considerably because each is
based on only five repeats. 
We therefore consider their distribution across the 50 noise realizations 
rather than interpreting individual values. 
We obtain $\operatorname{median}_i(R_{Z,i})=0.94$, close to the finite-sample
reference value of $0.92$. Aside from the single failed sampler
realization discussed above, we find no evidence that the uncertainties
reported by \dynesty{} systematically underestimate the observed
run-to-run variation.

Overall, these tests show that the adopted directed search follow-up
configuration is highly reproducible, while also illustrating a
limitation of nested sampling: successful termination of an
individual run does not by itself guarantee that all relevant
high-likelihood regions have been explored. 
In this study, one such failure was observed among the 250 runs.

These repeatability tests complement the preliminary \texttt{nlive} study 
presented in Appendix~\ref{sec:appendix-nlive-values}, whose primary purpose 
was to motivate the choice of the reference nested-sampling configurations 
adopted in this work. 
For the directed configuration, the observed stability of the recovered 
detection thresholds and efficiencies over the tested range of \texttt{nlive} values, 
together with the repeatability demonstrated here and the posterior calibration tests presented 
in Sec.~\ref{sec:parameter-estimation}, provides complementary empirical validation that 
the adopted nested-sampling configuration is sufficiently converged for this search setup. 

We have not performed a similar repeatability study for the all-sky 
follow-up due to its larger computational cost and because more 
detailed optimization for that scenario remains for future work,
likely including a hierarchical setup.

\subsection{Computational performance}
\label{sec:comp-cost}

The computational cost of stochastic \acs{cw} analyses depends primarily on 
the dimensionality of the parameter space and the sampler configuration. 
For the setups considered here, we find no significant dependence 
of the runtime on the injected signal strength. 
The following comparisons therefore focus on the effects of the analysis 
configuration and sampler choice.

For the directed follow-up, the reference \dynesty{} configuration adopted
in this work is computationally more expensive than the existing \ptemcee{}
implementation. The median wall-clock runtimes 
on the \acs{osg}/\acs{igwn} computing infrastructure
\cite{Bagnasco:2023gyj}
are $2.1\times10^3\,\mathrm{s}$ ($\sim35\,\mathrm{min}$) for \dynesty{} 
and $8.6\times10^2\,\mathrm{s}$ ($\sim14\,\mathrm{min}$) for \ptemcee{},
such that the \dynesty{} runs are approximately 2.4 times longer.

To investigate this difference, we additionally compare the number of
likelihood evaluations. 
For the reference \dynesty{} configuration, the
median number of likelihood evaluations is $2.2\times10^6$. 
For the
\ptemcee{} setup, with 3 temperatures, 300 walkers, and 1000 burn-in plus
1000 production steps, the nominal number of proposal likelihood evaluations
is $3\times300\times(1000+1000)=1.8\times10^6$; 
at each iteration, one proposal is generated for every walker at every
temperature, regardless of whether the proposal is subsequently accepted.
The number of likelihood evaluations for \dynesty{} is therefore only a
factor of approximately 1.2 larger than for \ptemcee{}, while its runtime
is a factor of approximately 2.4 larger. 
The difference in runtime is therefore not explained by the 
number of likelihood evaluations alone.
The median time per evaluation is 
$9.6\times10^{-4}\,\mathrm{s}$ for \dynesty{}, compared with 
$4.8\times10^{-4}\,\mathrm{s}$ for \ptemcee{},
indicating additional sampler- and 
implementation-specific overhead in the first.

Extending the analysis to include sky-position priors requires a larger
number of live points to recover the expected theoretical sensitivity and
is therefore considerably more computationally demanding. The median
wall-clock runtime for the all-sky follow-up \dynesty{} runs is
$5.1\times10^4\,\mathrm{s}$ ($\sim14\,\mathrm{h}$), compared with
$2.1\times10^3\,\mathrm{s}$ ($\sim35\,\mathrm{min}$) for the directed follow-up runs, 
corresponding to an increase by a factor of approximately 24.
The all-sky follow-up runtimes also show greater variation.

The median number of likelihood evaluations using \dynesty{} increases from
$2.2\times10^6$ for the runs in the directed configuration to 
$1.8\times10^7$ for the runs in the all-sky configuration,
corresponding to a factor of approximately 8.1. 
Thus, the increase in the
number of likelihood evaluations accounts for a substantial fraction, 
but not all, of the increase in runtime. 

These timing comparisons should not be interpreted as hardware-independent
sampler benchmarks. The analyses were performed on the heterogeneous
\acs{osg}/\acs{igwn} computing infrastructure and were not benchmarked on
identical compute nodes. Differences in the runtime per likelihood
evaluation may therefore reflect both sampler-specific overhead and
differences between the hardware used by individual jobs.

Despite the increased computational cost, the reference \dynesty{}
configuration recovers the expected sensitivity for the all-sky follow-up.
Among the single-stage \ptemcee{} configurations explored in this work,
none achieved comparable recovery efficiency. Moreover, the most
computationally demanding \ptemcee{} configurations tested for the all-sky
case were already slower than the reference \dynesty{} configuration,
while still recovering fewer signals. 
For both samplers, we expect hierarchical setups to be more efficient especially in this all-sky follow-up scenario.

The reference \dynesty{} configuration adopted throughout this 
work was chosen to prioritize robust recovery efficiency over 
computational performance.
Although Appendix~\ref{app:DYNESTY-configs} shows that alternative 
sampler settings can substantially reduce the runtime while 
maintaining comparable recovery efficiencies, a systematic optimization
of the nested-sampling configuration is beyond the scope of the present
work and is left for future investigation, along with an 
exploration of hierarchical multi-stage approaches.

\section{Parameter estimation and posterior validation}
\label{sec:parameter-estimation}

In addition to testing candidate signals at increased coherence time,
the stochastic samplers 
available through \bilby{} provide posterior samples for the 
searched parameters, enabling Bayesian \ac{pe}
and thus the astrophysical characterization of potential sources
for candidate \ac{cw} signals.
It is therefore important to verify 
the performance of the implemented framework in terms of
\ac{pe} precision and accuracy.

Previous work on stochastic sampling follow-ups of \ac{cw} candidates
has mostly focused on developing efficient strategies for
achieving sufficient signal recovery to exclude candidates 
or to pass them on to subsequent stages
\cite{ashtonHierarchicalMultistageMCMC2018, mirasolaComputationallyefficientFollowupPipeline2025, tenorioApplicationHierarchicalMCMC2021}, 
rather than on ensuring accurate parameter estimates in the final stage.
Exceptions include the \texttt{CWInPy} 
pipeline, whose \acs{pe}
results have been extensively tested for the known pulsar case
\cite{cwinpy},
but which has not been used for more general follow-ups of candidates
from directed or all-sky searches with the recent exception of the
G347.3$-$0.5 candidate follow-up
\cite{Mirasola:2026cpv}.
Furthermore, Ref.~\cite{covasNewFrameworkFollow2024}
demonstrated a comparison of \dynesty{} and \ptemcee{}
for a \acs{ligo} hardware injection,
and Ref.~\cite{martinsBayesianFrameworkFollowup2025}
provided a \ac{pp} test (see below)
for an all-sky follow-up setup with \dynesty{}.

We focus here on a well-established test of
\Acs{pe} accuracy, namely that the resulting posterior 
distributions are well calibrated.
To this end, we perform a \ac{pp} test
\cite{Cook:2006zir}, 
computing, for each injection, the percentile rank of the true value of each parameter within its recovered posterior.
For well-calibrated posteriors, these percentile ranks should be uniformly distributed.
Equivalently, across repeated injections, the fraction of true parameter values contained within a given credible interval should match its nominal credible level.

\begin{figure*}[t]
    \centering
    \begin{minipage}{0.48\textwidth}
        \centering
        \includegraphics[width=\linewidth]{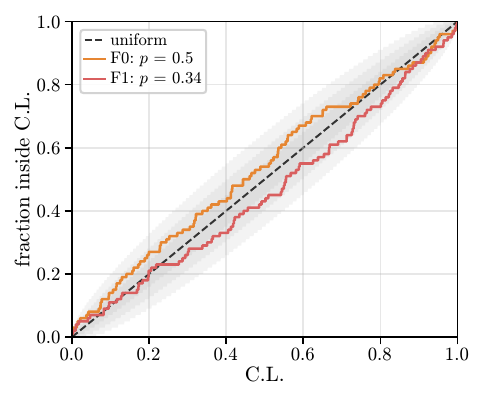}
    \end{minipage}
    \hfill
    \begin{minipage}{0.48\textwidth}
        \centering
        \includegraphics[width=\linewidth]{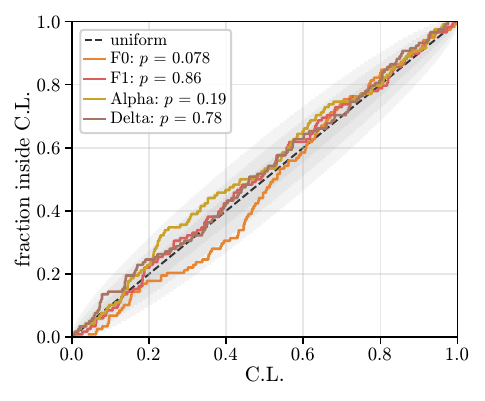}
    \end{minipage}
    \caption{        
    \Acs{pp} plots obtained from 100 injections at depth $D=125$ using the \dynesty{} sampler for the directed follow-up (left) and the all-sky follow-up (right). The shaded regions show the expected 1-, 2- and 3-$\sigma$ statistical fluctuations for perfectly calibrated posteriors. The Fisher-combined $p$-values over the plotted parameters are 0.47 for the directed follow-up and 0.82 for the all-sky follow-up.
    }
    \label{fig:pp-plots}
\end{figure*}

As seen in Fig.~\ref{fig:pp-plots},
the \acs{pp} tests show no statistically significant deviations from the expected
uniform distribution for either the directed or the all-sky
follow-up configuration with \dynesty{}. 
For each parameter, we quantify the
agreement with the expected uniform distribution using a one-sample
Kolmogorov--Smirnov test \cite{masseyKolmogorovSmirnovTestGoodness1951} 
on the recovered percentile ranks. 
We then combine the parameter-level $p$-values using Fisher's method
\cite{fisherStatisticalMethodsResearch1992},
which maps the individual $p$-values to a $\chi^2$ distribution to evaluate 
their joint significance. 
This yields final Fisher-combined $p$-values of 
0.47 for the directed search follow-up and 0.82 for the all-sky follow-up. 
These values indicate that the observed deviations from the diagonal 
are consistent with the statistical fluctuations expected 
for well-calibrated posteriors.
We therefore find no evidence that the recovered credible intervals 
are miscalibrated in either configuration.

\begin{figure}[tbp]
    \centering
    \includegraphics[width=\columnwidth]{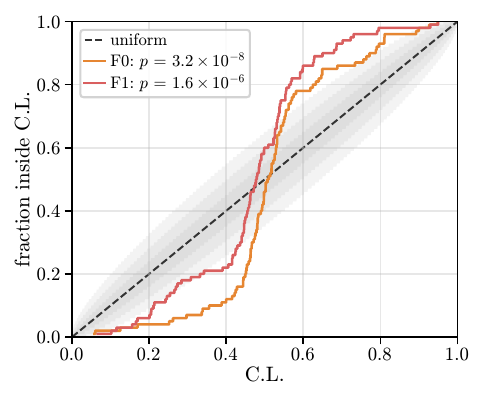}
    \caption{
    \Acs{pp} plot obtained from 100 injections at depth $D=100$ using the \ptemcee{} sampler for the directed follow-up.
    The shaded regions show the expected 1-, 2- and 3-\(\sigma\) statistical fluctuations for perfectly calibrated posteriors.
    The Fisher-combined $p$-value is $1.6\times10^{-12}$.}
    \label{fig:fixedsky_pp_pyfstat}
\end{figure}

By contrast, the fixed-sky \acs{pp} plot obtained with the \ptemcee{}
configuration considered here shows
systematic deviations from the expected uniform distribution,
as can be seen in Fig.~\ref{fig:fixedsky_pp_pyfstat}. 
The Fisher-combined $p$-value for the plotted parameters is $1.6\times10^{-12}$,
indicating that its posterior credible intervals are not
well calibrated under the conditions of this study.
This also demonstrates that achieving the 
expected theoretical detection efficiency alone
is not a sufficient criterion for a setup that will also provide
reliable \ac{pe}.

Together with the detection-efficiency results of Sec.~\ref{sec:results}, 
these tests show that the integration of \bilby{} with \pyfstat{} can enable both 
candidate recovery and \ac{pe}.
In addition, the tested \dynesty{} configurations, albeit expensive,
provide both the expected detection performance 
and well-calibrated \ac{pe}.

Since 90\% credible intervals are commonly reported for \acs{pe} results in \acs{gw} astronomy, 
we also show centered-offset diagnostics for the 90\% credible intervals in
Appendix~\ref{sec:appendix-coverage}.

\section{Conclusions}
\label{sec:conclusions}
We have presented the integration of the \bilby{} inference library
\cite{ashtonBilbyUserfriendlyBayesian2019}
into \pyfstat{}
\cite{keitelPyFstatPythonPackage2021,ashtonPyFstatPyFstatV2302025}, 
enabling the use of the wide variety of samplers available through
\bilby{} within the existing \pyfstat{} search infrastructure
for \ac{cw} analyses.
The integration preserves the existing search classes, 
signal-model implementation, and detection-statistic calculations 
while replacing only the stochastic sampling engine.

In terms of functionality, the interface presented here is comparable 
to other recently proposed frameworks that combine $\mathcal{F}$-statistic-based analyses 
with the samplers available through \bilby{} 
\cite{covasNewFrameworkFollow2024,martinsBayesianFrameworkFollowup2025}. 
Unlike those frameworks, which are not publicly available, 
the present implementation is integrated directly into the open-source \pyfstat{} codebase, 
making these capabilities readily accessible to the broader \acs{cw} community. 
We also provide the first systematic validation 
of parameter-estimation performance with \pyfstat{}, 
complementing previous studies that focused primarily 
on candidate recovery and detection performance.

The proposed framework has been validated using injection studies 
in single-stage fully coherent
directed and all-sky follow-up configurations. 
In both cases, the measured
detection efficiencies are consistent with the theoretical sensitivity
predictions,
demonstrating that the expected detection performance is recovered.
Furthermore, \acs{pp} tests show no
evidence of posterior miscalibration, indicating that the inferred
credible intervals provide statistically consistent uncertainty
estimates for the recovered source parameters.

As expected, allowing the sky position to vary increases both the computational cost and the detection threshold through the larger parameter space. 
Nevertheless, the reference \dynesty{} nested-sampling configuration
considered here continues to recover the predicted sensitivity. 
Among the single-stage \ptemcee{} configurations explored in this work, 
we did not identify one that achieved comparable recovery performance for the
all-sky follow-up considered here. 
While this does not preclude alternative \acs{mcmc} strategies, 
including hierarchical multistage follow-up schemes
\cite{ashtonHierarchicalMultistageMCMC2018,tenorioApplicationHierarchicalMCMC2021,mirasolaComputationallyefficientFollowupPipeline2025},
these results demonstrate that nested sampling can be used
for single-stage coherent follow-up analyses in higher-dimensional parameter spaces.

The interface supports two complementary applications: 
\ac{pe} and coherent candidate follow-up. 
While \ac{pe} analyses require well-converged posterior distributions 
and well-calibrated credible intervals, 
follow-up analyses primarily require robust candidate recovery and localization.
These objectives place different demands on the stochastic sampler, 
suggesting that sampler configurations should ultimately be chosen according 
to the intended application rather than relying on a single default setup. 
A systematic exploration of such application-specific tuning, however, 
lies beyond the scope of the present work.

Accordingly, the sampler settings adopted in this work were selected 
to provide reliable performance across the follow-up configurations considered, 
rather than to optimize computational efficiency for a particular application. 
Future work will investigate application-specific nested-sampling strategies, 
including alternative stopping criteria and sampling methods, 
with the aim of reducing the computational cost while maintaining 
the performance required for \ac{pe} and candidate follow-up.

Overall, these results demonstrate that the integration presented here provides
a reliable and flexible Bayesian framework for coherent \ac{cw} searches
while extending the open-source package \pyfstat{} with access to a broad
range of modern stochastic samplers through \bilby{}.

\section{Acknowledgements}

We thank Greg Ashton for helpful comments on the manuscript 
and Lorenzo Mirasola for useful discussions on the \texttt{cows3} sensitivity-estimation framework.

This work was supported by the yearly plan of the Tourist Stay Tax for 2023 (ITS2023-086 - Programa de Fomento de la Investigación), the Universitat de les Illes Balears (UIB) with funds from the Programa de Foment de la Recerca i la Innovació de la UIB 2024-2026 (supported by the yearly plan of the Tourist Stay Tax ITS2023-086); the Spanish Agencia Estatal de Investigación grants PID2022-138626NB-I00, RED2024-153978-E, RED2024-153735-E, funded by MICIU/AEI/10.13039/501100011033 and the ERDF/EU; and the Comunitat Autònoma de les Illes Balears through an ``Ajut per a projectes de recerca científica i tecnològica'' (PRD2025\_100) and through
the Conselleria d'Educació i Universitats with funds from the ERDF (SINCO2022/18146 - Plataforma HiTech-IAC3-BIO);
and by COST action SCALES CA24139, supported by COST (European Cooperation in Science and Technology). 
This material is based upon work supported by NSF's LIGO Laboratory, which is a major facility fully funded by the National Science Foundation.
The authors are grateful for computational resources provided by the LIGO Laboratory and supported by National Science Foundation Grants PHY-0757058 and PHY-0823459.

This document has been assigned LIGO document number \href{https://dcc.ligo.org/LIGO-P2600436}{P2600436}.

\appendix

\section{Noise likelihood derivation}
\label{sec:noise-likelihood}
This appendix derives the Gaussian-noise likelihood entering Eq.~\eqref{eq:logL}. 
As discussed in Sec.~\ref{sec:bilby-integration}, this term provides 
the data-dependent normalization required for Bayesian evidence calculations.

Under the Gaussian-noise hypothesis, $\mathcal{H}_\mathrm{N}$, the
likelihood of the data $x$ is
\begin{equation}
    \mathcal{L}_\mathrm{N}(x)
    = P(x|\mathcal{H}_\mathrm{N})
    = \kappa\exp\left[-\frac{1}{2}(x|x)\right],
    \label{eq:L_noise_appendix}
\end{equation}
where $(x|x)$ denotes the standard noise-weighted inner product 
and $\kappa$ is the normalization constant.

Following the Gaussian-noise formalism described in 
Ref.~\cite{prixFstatisticImplementation2018}, the inner product can be written as
\begin{equation}
    (x|x)
    = 4\operatorname{Re}\int_0^\infty
    \sum_{X,Y}
    \widetilde{x}^{X*}(f)
    \left[S^{-1}(f)\right]_{XY}
    \widetilde{x}^{Y}(f)\,df.
\end{equation}

For uncorrelated detector noise, the cross-spectral-density matrix $S^{XY}(f)$
is diagonal, which gives
\begin{equation}
    (x|x)
    = \sum_{X=1}^{N_\mathrm{Det}}(x^X|x^X)
    = 4\sum_{X=1}^{N_\mathrm{Det}}
    \int_0^\infty
    \frac{|\widetilde{x}^{X}(f)|^2}{S^{X}(f)}\,df,
    \label{eq:scalar-product-x}
\end{equation}
where $N_\mathrm{Det}$ is the number of detectors
and $S^{X}(f)$ is the one-sided noise  power spectral density of detector $X$.
In practice, this quantity is not known \emph{a priori} and must be estimated from the data. 
In the implementation presented here, 
we estimate $S^{X}(f)$ 
using the standard running-median procedure implemented in \texttt{LALSuite}
\cite{lalsuite, swiglal}.
We denote the resulting estimate for detector $X$, \ac{sft}
$\alpha$, and frequency bin $k$ by $\widehat{S}^{X}_{\alpha k}$.

Discretizing Eq.~\eqref{eq:scalar-product-x} and substituting
$S^{X}(f)$ by its estimate $\widehat{S}^{X}_{\alpha k}$
gives
\begin{equation}
    (x|x)
    \simeq
    4\sum_{X,\alpha,k}
    \frac{|\widetilde{x}^{X}_{\alpha k}|^2\Delta f}
         {\widehat{S}^{X}_{\alpha k}}
    =
    \frac{4}{T_\mathrm{SFT}}
    \sum_{X,\alpha,k}
    \frac{|\widetilde{x}^{X}_{\alpha k}|^2}
         {\widehat{S}^{X}_{\alpha k}},
\end{equation}
where $\Delta f = 1/T_\mathrm{SFT}$ is the \acs{sft} frequency resolution,
$T_\mathrm{SFT}$ is the duration of each \acl{sft},
$\alpha$ labels the \acp{sft}, and $k$ labels the frequency bins.

To determine the normalization constant, $\kappa$, we consider the statistical 
distribution of the \acs{sft} Fourier coefficients. 
Under the Gaussian-noise hypothesis, each positive-frequency \acs{sft} coefficient 
can be written as
\begin{equation} 
    \widetilde{x}^{X}_{\alpha k} = a^{X}_{\alpha k} + i\,b^{X}_{\alpha k} ,
\end{equation} 
where $a^{X}_{\alpha k}$ and $b^{X}_{\alpha k}$ are independent
zero-mean Gaussian random variables with equal variance,
\begin{equation}
\left(\sigma_{\alpha k}^{X}\right)^2
=
\operatorname{Var}(a^{X}_{\alpha k})
=
\operatorname{Var}(b^{X}_{\alpha k})
=
\frac{T_\mathrm{SFT}}{4}
\widehat{S}^{X}_{\alpha k}.
\end{equation}

The joint normalization of the corresponding two-dimensional Gaussian distribution 
is therefore $1/(2\pi\left(\sigma_{\alpha k}^{X}\right)^2)$. 
Since the positive-frequency \acs{sft} Fourier coefficients are assumed to be statistically
independent, 
the normalization constant factorizes into the product
\begin{equation} 
    \kappa = \prod_{X,\alpha,k} \frac{1} {2\pi\left(\sigma_{\alpha k}^{X}\right)^2} 
    = \prod_{X,\alpha,k} \frac{2} {\pi T_\mathrm{SFT}\widehat{S}^{X}_{\alpha k}}. 
\end{equation} 

Substituting the expressions for $\kappa$ and $(x|x)$ 
into Eq.~\eqref{eq:L_noise} yields the final expression for the Gaussian-noise likelihood,
\begin{equation}
\mathcal{L}_\mathrm{N}
=
\prod_{X,\alpha,k}
\frac{2}
{\pi T_\mathrm{SFT}\widehat{S}^{X}_{\alpha k}}
\exp\left[
-
\frac{2
|\widetilde{x}^{X}_{\alpha k}|^2}
{T_\mathrm{SFT}\widehat{S}^{X}_{\alpha k}}
\right].
\end{equation}

\section{Theoretical sensitivity estimates}
\label{sec:appendix-theoretical-estimate}

This appendix describes the two theoretical sensitivity estimates used to
predict the detection efficiency of the coherent follow-ups of the main text. 
Both calculations are based on the non-central $\chi^2$ distribution of the
$\mathcal{F}$-statistic, but differ in how they average the signal-to-noise
ratio over the source population and detector response.

Under the signal hypothesis, the coherent detection statistic
$2\mathcal{F}$ follows a non-central $\chi^2$ distribution with four
degrees of freedom and non-centrality parameter $\rho^2$. The detection
probability for a fixed threshold $2\mathcal{F}_{\rm th}$ is therefore
\begin{equation}
p_\mathrm{det}
=
P\left(2\mathcal{F}>2\mathcal{F}_{\rm th};\rho^2\right),
\label{eq:p-det}
\end{equation}
which is evaluated using the survival function of the non-central
$\chi^2_4(\rho^2)$ distribution. Because $2\mathcal{F}$ varies between noise
realizations, a signal can remain below the threshold even when the search
identifies the correct signal mode. This statistical loss is included in the
theoretical efficiency curves; a measured efficiency below these predictions
can therefore reveal additional losses introduced by the search procedure.

\subsection{Semi-analytic estimate}

The semi-analytic calculation follows
Ref.~\cite{wetteEstimatingSensitivityWideparameterspace2012}.
Estimating a representative value of $\rho^2$ for the full injection set to then evaluate \ref{eq:p-det}
requires averaging over the detector response, which depends on the source
sky position and orientation.

For fixed source parameters, the squared signal-to-noise ratio of a signal with amplitude $h_0$
observed for a time $T_\mathrm{s}$ in noise with one-sided power spectral density $S_n$
is given by
\begin{equation}
\rho^2(\iota,\alpha,\delta,\psi)
=
\frac{h_0^2 T_\mathrm{s}}{S_n}
\left[
a_+^2(\iota)\left\langle F_+^2 \right\rangle_t
+
a_\times^2(\iota)\left\langle F_\times^2 \right\rangle_t
\right],
\label{eq:rho2}
\end{equation}
as shown in Ref.~\cite{wetteEstimatingSensitivityWideparameterspace2012}. 
In Eq.~\ref{eq:rho2}, the angle brackets denote an average over the
observation time for fixed sky position and polarization angle. The functions
$F_+(t; \alpha, \delta, \psi)$ and $F_\times(t; \alpha, \delta, \psi)$ are the detector antenna pattern functions, and
$a_+$ and $a_\times$ are the amplitude modulation coefficients, which depend only on $\cos\iota$ and are given by
\begin{equation}
a_+(\iota)=\frac{1}{2}(1+\cos^2\iota),
\qquad
a_\times(\iota)=\cos\iota .
\end{equation}

To obtain a simple analytic prediction, we replace the antenna-pattern functions by their averages over sky position, polarization angle, and observation time for interferometers with orthogonal arms
(see also Ref.~\cite{prixFstatisticImplementation2018}):
\begin{equation}
\left\langle F_+^2 \right\rangle_{\alpha,\sin\delta,\psi,t}
=
\left\langle F_\times^2 \right\rangle_{\alpha,\sin\delta,\psi,t}
=
\frac{1}{5}.
\end{equation}

This approximation averages the detector response over sky position,
polarization, and time, but keeps the inclination dependence explicit through
$a_+(\iota)$ and $a_\times(\iota)$. It is therefore not a full population average
over all angular variables. 

For an isotropically oriented source population, we assume
$\cos\iota \sim \mathcal{U}(-1,1)$. The population-averaged detection
efficiency is then obtained by averaging the detection probability over the
inclination distribution,
\begin{equation}
\langle p_{\rm det} \rangle
=
\int_{-1}^{1}
p_{\rm det}\!\left[\rho^2(\cos\iota)\right]
\,\frac{d(\cos\iota)}{2},
\end{equation}
where $\rho^2(\cos\iota)$ is computed from Eq.~\ref{eq:rho2}. 
Because the detection probability depends
nonlinearly on the signal-to-noise ratio, the inclination average must be
performed over $p_{\rm det}$ itself rather than over $\rho^2$. The integral is
evaluated numerically by Monte Carlo sampling of $\cos\iota$.

This semi-analytic approximation provides a simple and computationally
inexpensive prediction for the detection efficiency while retaining the
dominant dependence on the source inclination. It serves as a reference
against which the measured detection efficiencies from the injection
studies can be compared.

\subsection{The \texttt{cows3} estimate}

The second estimate is obtained with the open-source package \texttt{cows3}
\cite{mirasolaComputationallyefficientFollowupPipeline2025}, which implements
the sensitivity-estimation framework of
\cite{dreissigackerFastAccurateSensitivity2018}. 
Rather than replacing the
antenna-pattern factors with their population averages, \texttt{cows3} samples
the source and detector parameters to construct the distribution of $\rho^2$
across the source population. The detection probability is evaluated for the
sampled values of $\rho^2$ and then averaged numerically, providing a more
complete treatment of the source population and detector response.

\section{Tests of injection recovery with different configurations}
\label{app:additional-injection-recovery-studies}

This appendix collects additional explorations of the injection-recovery
performance.  We assess the robustness of the results to the number of live
points, the nested-sampling walk strategy, and the use of a realistic observing
schedule.

\subsection{Preliminary exploration of \texttt{nlive} values}
\label{sec:appendix-nlive-values}

\begin{figure*}[tbp]
    \centering
    \begin{minipage}{0.48\textwidth}
        \centering
        \includegraphics[width=\columnwidth]{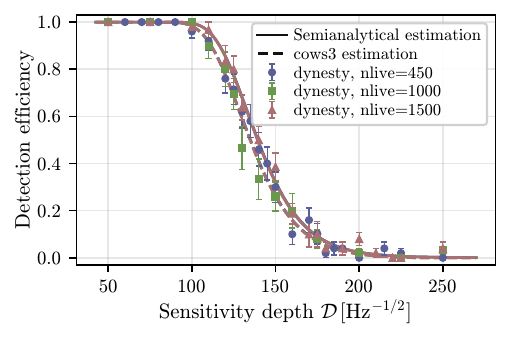}
    \end{minipage}
    \hfill
    \begin{minipage}{0.48\textwidth}
        \centering
        \includegraphics[width=\columnwidth]{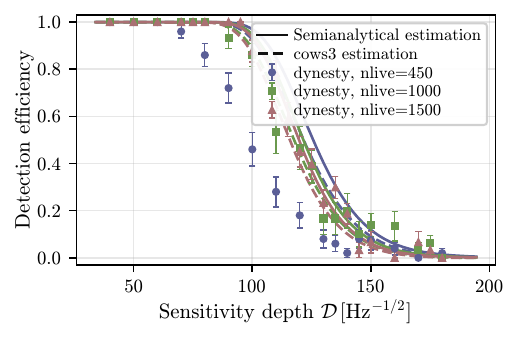}
    \end{minipage}
    \caption{Detection efficiency as a function of sensitivity depth for different
    values of \texttt{nlive} for the directed follow-up (left) and
    the all-sky follow-up (right). Markers indicate the measured recovery
    fractions from fixed-orientation injection studies, with error bars showing
    the binomial statistical uncertainty. The curves show the corresponding
    semi-analytic and \texttt{cows3} sensitivity estimates for each configuration.
    This comparison is used to select the reference \texttt{nlive} values adopted
    in the main text.
    }
    \label{fig:efficiencies_nlive_comparison}
\end{figure*}

The number of live points is one of the main parameters controlling the performance 
of nested-sampling algorithms, affecting both the computational cost and the 
accuracy with which the parameter space is explored. 
This subsection describes the preliminary studies used to select the reference 
values of \texttt{nlive} adopted in the main text.
Although these studies were primarily performed to determine suitable nested-sampling
configurations, they also provide an empirical assessment of the stability of the 
recovered results with respect to the number of live points.

For simplicity, these studies were performed using contiguous 100-day
datasets with fixed-orientation injections,
with $\cos\iota=0, \ \Psi=0$. 
Since the source orientation mainly affects the absolute
detection efficiency rather than the behavior of the nested sampler, the
qualitative dependence on \texttt{nlive} is expected to be similar for the
isotropic-orientation studies presented in the main text.

Fig.~\ref{fig:efficiencies_nlive_comparison} compares the measured detection 
efficiencies obtained with \texttt{nlive} = 450, 1000, and 1500 for the directed 
and all-sky follow-up configurations.
The semi-analytic and \texttt{cows3} estimates differ slightly in this
comparison. The \texttt{cows3} result should be regarded as the more accurate
prediction because it explicitly samples the relevant source and detector
parameters, whereas the semi-analytic calculation uses population-averaged
antenna-pattern factors.

For the fixed-sky configuration, all tested values of \texttt{nlive} yield statistically consistent detection efficiencies that agree with the theoretical sensitivity estimates. Increasing \texttt{nlive} does not produce a measurable improvement, indicating that \texttt{nlive}=450 provides sufficient parameter-space
exploration in this case.

The analysis with sky-position priors exhibits a stronger dependence on \texttt{nlive}. 
In contrast to the fixed-sky case, the configurations with 
$\texttt{nlive}=450$ and $\texttt{nlive}=1000$ do not recover the 
theoretical detection efficiency over the transition region of the curve, 
indicating that the higher-dimensional parameter space is not explored 
sufficiently with these settings. Increasing the number of live points 
improves the recovery efficiency, with $\texttt{nlive}=1500$ providing 
the first tested configuration consistent with the theoretical sensitivity 
estimates. These results motivate the choice of $\texttt{nlive}=1500$ as 
the reference configuration for the all-sky analyses presented in the main text.

\subsection{Additional \dynesty{} configurations}
\label{app:DYNESTY-configs}

\begin{figure*}[]
    \centering
    \begin{minipage}{0.48\textwidth}
        \centering
        \includegraphics[width=\columnwidth]{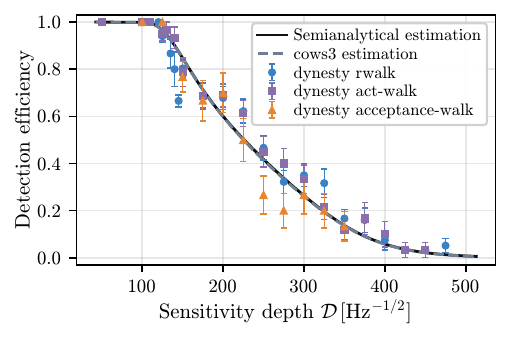}
    \end{minipage}
    \hfill
    \begin{minipage}{0.48\textwidth}
        \centering
        \includegraphics[width=\columnwidth]{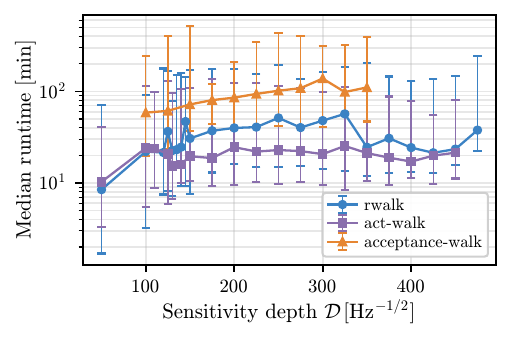}
    \end{minipage} 
    \caption{Comparison of different \dynesty{} walk strategies for the
    directed follow-up. Left: detection efficiency as a function of
    sensitivity depth for the \texttt{acceptance-walk}, \texttt{act-walk}, and
    \texttt{rwalk} sampling methods. Right: median runtime as a function of
    sensitivity depth for the same configurations. Error bars show the full
    range of measured runtimes at each depth, from the fastest to the slowest
    completed run. The \texttt{act-walk} and
    \texttt{rwalk} configurations recover detection efficiencies consistent with
    the theoretical predictions, while \texttt{acceptance-walk} shows a modest
    reduction in recovery efficiency.}
    \label{fig:dynesty_walkers_comparison}
\end{figure*}

To assess the impact of the nested-sampling walk strategy on the search
performance, we performed a series of additional fixed-sky injection
studies using contiguous 100-day datasets with isotropic orientation. 
The comparison focuses on the \dynesty{} walk strategies, 
while keeping the remaining sampler settings fixed.

Fig.~\ref{fig:dynesty_walkers_comparison} compares the detection
efficiency and runtime obtained with the
\texttt{acceptance-walk}, \texttt{act-walk}, and \texttt{rwalk} 
sampling methods of \dynesty{} for the fixed-sky configuration.
All configurations use \texttt{nlive}=450, an evidence tolerance of
\texttt{dlogz}=0.1, and the \texttt{live} bounding method. For the
walk-specific settings, \texttt{rwalk} uses \texttt{walks}=100,
\texttt{act-walk} uses \texttt{nact}=1 and \texttt{maxmcmc}=100, and
\texttt{acceptance-walk} uses \texttt{naccept}=20,
\texttt{walks}=100, and \texttt{maxmcmc}=5000.

The walk strategy is only one of several parameters that affect the
performance of \dynesty{}.
Therefore, the results presented here do
not constitute an exhaustive optimization of the sampler.

The \texttt{act-walk} and \texttt{rwalk} configurations recover detection
efficiencies consistent with the theoretical sensitivity estimates within
the statistical uncertainties of the injection study. The
\texttt{acceptance-walk} configuration, however, shows a modest reduction
in recovery efficiency over part of the transition region, while also
requiring substantially longer runtimes. 
Consequently, fewer injections were performed for this configuration than for the others.

The principal distinction between the remaining well-performing strategies
is their computational cost: \texttt{act-walk} achieves the shortest median
runtime, while \texttt{rwalk} is consistently slower but still recovers the
expected detection efficiency.

The main analyses presented in this paper were carried out using the
\texttt{rwalk} strategy, since they were completed before this
comparison of walk strategies was performed. Given the relatively modest
runtime reduction obtained with \texttt{act-walk}, these analyses were
not repeated.

\subsection{Results using O4a-timestamp datasets}
\label{app:o4timestamps}

\begin{figure*}[]
    \centering
    \begin{minipage}{0.48\textwidth}
        \centering
        \includegraphics[width=\linewidth]{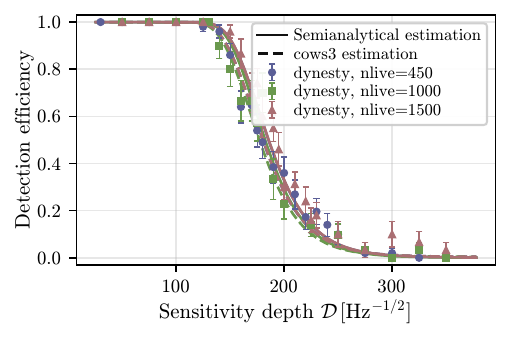}
    \end{minipage}
    \hfill
    \begin{minipage}{0.48\textwidth}
        \centering
        \includegraphics[width=\linewidth]{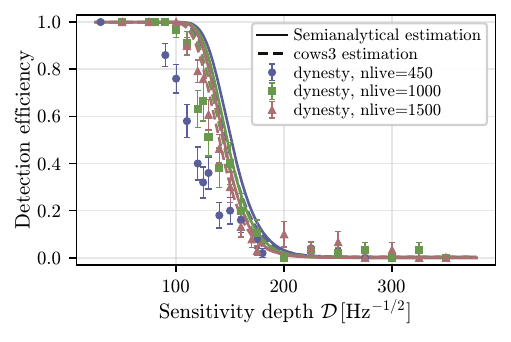}
    \end{minipage}
    \caption{
    Detection efficiency as a function of sensitivity depth for the
    O4a-timestamp Gaussian-noise datasets, for the fixed-sky (left) and
    all-sky (right) follow-ups.
    Markers indicate the measured detection efficiencies for
    $\texttt{nlive}=450$, $1000$, and $1500$, while the solid and dashed
    curves show the corresponding semi-analytic and \texttt{cows3}
    sensitivity estimates, respectively.}
    \label{fig:effic_o4_timestamps}
\end{figure*}

To assess the robustness of the results with respect to the observing
schedule, we repeated the preliminary \texttt{nlive} studies using
Gaussian-noise datasets constructed from detector timestamps 
from the first part of the fourth \acs{lvk} observing run (O4a)
\cite{LIGOScientific:2025snk}.
As in Appendix~\ref{sec:appendix-nlive-values}, these
exploratory studies use fixed-orientation injections.

The injected parameter values and the centers of the search priors 
are the same as in the contiguous 100-day fixed-orientation study.  
The frequency and spin-down priors are uniform, centered on the corresponding injected values, 
with widths
$\Delta f=2\times10^{-5}\,\mathrm{Hz}$ and
$\Delta \dot f=2\times10^{-12}\,\mathrm{Hz\,s^{-1}}$.
For the all-sky follow-ups, the sky-position priors are uniform
and centered on the injected sky position and have widths
$\Delta\alpha=\Delta\delta=0.05\,\mathrm{rad}$.
Thus, relative to the contiguous 100-day datasets, the frequency and sky-position widths
are unchanged, while the spin-down width is narrower than the
$\Delta \dot f=1.8\times10^{-10}\,\mathrm{Hz\,s^{-1}}$ used there.
The tested values of \texttt{nlive} and all other \dynesty{} settings
are the same as in the contiguous 100-day study.

Fig.~\ref{fig:effic_o4_timestamps} compares the detection efficiencies
obtained with different values of \texttt{nlive} for the directed search follow-up
and for the all-sky search follow-up.

For the directed configuration, all tested values of \texttt{nlive}
recover detection efficiencies consistent with the theoretical
sensitivity estimates, confirming the behavior observed for the
contiguous 100-day datasets. As before, \texttt{nlive}=450 is
sufficient to recover the expected sensitivity, with no measurable
improvement obtained by increasing the number of live points.

The all-sky follow-up configuration again exhibits a stronger dependence on
\texttt{nlive}. Increasing the number of live points improves the
agreement with the theoretical sensitivity estimates, as found for the
contiguous 100-day datasets. However, even the largest value considered
here, \texttt{nlive}=1500,
greater deviations from the theoretical
predictions than for the corresponding contiguous 100-day datasets.
This suggests that the more irregular O4a observing schedule 
may require a more thorough exploration of the higher-dimensional parameter space,
potentially motivating additional
sampler tuning or larger values of \texttt{nlive}.

Overall, these results reproduce the qualitative trends observed 
for the contiguous 100-day datasets, 
although the all-sky follow-up appears to require a more demanding
sampler configuration.

\section{Additional posterior-calibration diagnostics}
\label{sec:appendix-coverage}
As an additional check of the \ac{pe} results discussed in
Sec.~\ref{sec:parameter-estimation}, we examine the position of each injected
value within the corresponding 90\% credible interval.
We use equal-tailed intervals, bounded by the 5th and 95th posterior
percentiles, $q_{5}$ and $q_{95}$, respectively, and define the normalized
centered offset
\begin{equation}
z_{90}
=
\frac{(q_{95}+q_{5})/2-\theta_{\mathrm{inj}}}
     {(q_{95}-q_{5})/2},
\label{eq:normalized-posterior-offset}
\end{equation}
where $\theta_{\mathrm{inj}}$ is the injected value of the parameter of interest. 
Thus, $z_{90}=0$ means that the injected value coincides with the midpoint of
the credible interval, while $|z_{90}|\leq1$ means that the injected value lies
inside that interval.  For well-calibrated posteriors, these distributions
should show no systematic displacement from zero and approximately 90\% of
the injections should lie between $-1$ and $1$.

\begin{figure}[]
    \centering
    \includegraphics[width=\columnwidth]{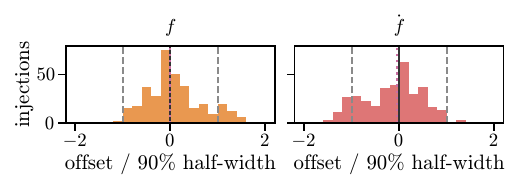}
    \caption{Normalized centered offsets, as defined in
    Eq.~\eqref{eq:normalized-posterior-offset}, for the \dynesty{}
    directed follow-up at $\mathcal{D}=145$.  The redsolid vertical lines
    mark zero, the gray dashed lines mark the bounds $z_{90}=\pm1$, and the blak dotted
    lines mark the median offsets.  The empirical coverages of the nominal
    90\% credible intervals are 0.87 for $f$ and 0.86 for $\dot f$.}
    \label{fig:normalized-offsets-fixed}
\end{figure}

\begin{figure}[]
    \centering
    \includegraphics[width=\columnwidth]{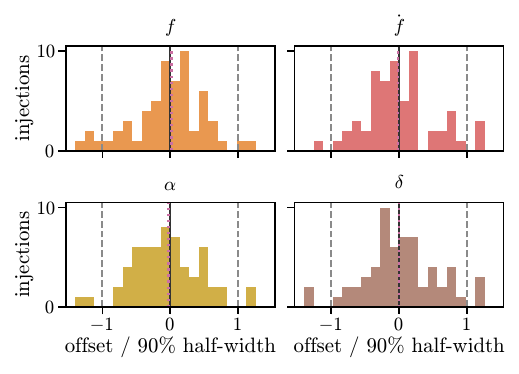}
    \caption{Normalized centered offsets, as defined in
    Eq.~\eqref{eq:normalized-posterior-offset}, for the \dynesty{}
    all-sky follow-up at $\mathcal{D}=120$.  The red solid
    vertical lines mark zero, the gray dashed lines mark the bounds
    $z_{90}=\pm1$, and the blackdotted lines mark the median offsets.  The empirical
    coverages of the nominal 90\% credible intervals are 0.90 for $f$, 0.93 for
    $\dot f$, 0.93 for $\alpha$, and 0.92 for $\delta$.}
    \label{fig:normalized-offsets-allsky}
\end{figure}

Figs.~\ref{fig:normalized-offsets-fixed}
and~\ref{fig:normalized-offsets-allsky} show these diagnostics for the sampled
sensitivity depth with measured detection efficiency closest to 90\% in each
configuration: 
$\mathcal{D}=145$ for the directed follow-up and
$\mathcal{D}=120$ for the all-sky follow-up.  
For the directed follow-up, the empirical coverages of the nominal 90\%
credible intervals are 0.87 for $f$ and 0.86 for $\dot f$.
For the all-sky follow-up, the corresponding coverages are 0.90 for $f$,
0.93 for $\dot f$, 0.93 for $\alpha$, and 0.92 for $\delta$.
The coverage fractions are therefore close to their nominal value, and the
offset distributions are centered near zero, consistent with the \ac{pp}
tests in Sec.~\ref{sec:parameter-estimation}.

We also find that the coverage remains broadly consistent with 90\% across the range of sampled depths in both configurations.

\clearpage

\bibliographystyle{apsrev-doi}
\bibliography{citations_zotero,bibliography}

\end{document}